# The LOFT mission: new perspectives in the research field of (accreting) compact objects

E. Bozzo[1], L. Stella[2], M. van der Klis[3], A. Watts[3], D. Barret[4], J. Wilms[5], P. Uttley[3], J. W. den Herder[6], M. Feroci[7] **(on behalf of the LOFT Consortium)**

[1]ISDC, department of Astronomy, University of Geneva, Chemin d'Ecogia 16, CH-1290 Versoix, Switzerland
[2]INAF - Osservatorio Astronomico di Roma, Via Frascati 33, I-00044 Roma, Italy
[3]Astronomical Institute Anton Pannekoek, University of Amsterdam, Science Park 904, 1090GE Amsterdam, The Netherlands
[4]Universitè de Toulouse, UPS-OMP, IRAP, Toulouse, France & CNRS, Institut de Recherche en Astrophysique et Planetologie, 9 Av. colonel Roche, BP 44346, F-31028 Toulouse cedex 4, France
[5]Dr. Karl-Remeis-Sternwarte and ECAP, Stemwartstr. 7, 96049 Bamberg, Germany
[6]SRON Netherlands Institute for Space Research, Sorbonnelaan 2, 3584 CA Utrecht, Netherlands
[7]INAF-IASF, Via del Fosso del Cavaliere 100, I-00133 Roma, Italy

**Abstract.** LOFT, the Large Observatory For X-ray Timing, is one of five ESA M3 candidate missions. It will address the Cosmic Vision theme: "Matter under Extreme Conditions". By coupling for the first time a huge collecting area for the detection of X-ray photons with CCD-quality spectral resolution (15 times bigger in area than any previously flown X-ray instrument and >100 times bigger for spectroscopy than any similar-resolution instrument), the instruments onboard LOFT have been designed to (i) determine the properties of ultradense matter by reconstructing its Equation of State through neutron star mass and radius measurements of unprecedented accuracy; (ii) measure General Relativity effects in the strong field regime in the stationary spacetimes of neutron stars and black holes of all masses down to a few gravitational radii. Besides the above two themes, LOFT's observations will be devoted to "observatory science", providing new insights in a number of research fields in high energy astrophysics (e.g. Gamma-ray Bursts). The assessment study phase of LOFT, which ended in September 2013, demonstrated that the mission is low risk and the required Technology Readiness Level can be easily reached in time for a launch by the end of 2022.

## 1 Introduction

The LOFT mission [1] has been primarily designed to address two fundamental questions that are part of ESA's Cosmic Vision Theme "Matter under extreme conditions":

- What is the equation of state of ultra-dense matter in neutron stars?
- Does matter orbiting close to the event horizon follow the predictions of general relativity?

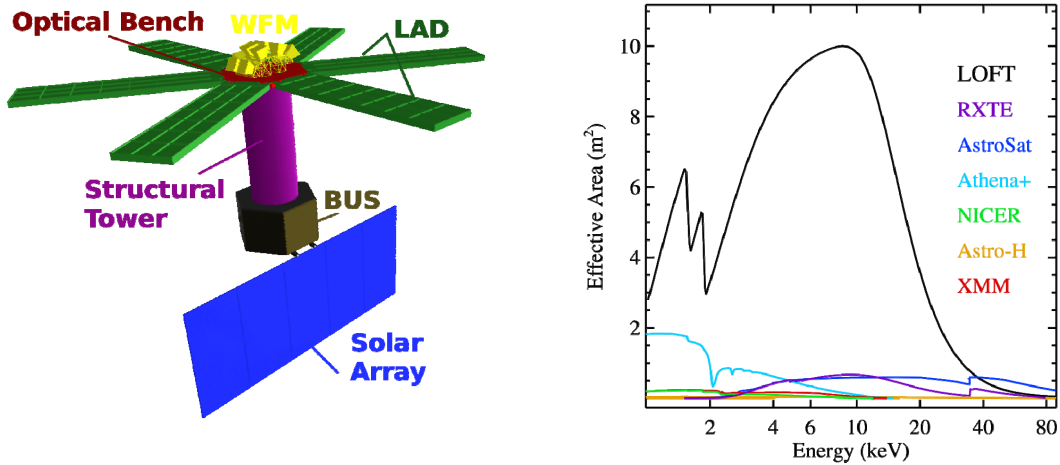

**Figure 1. Left:** a sketch of the LOFT satellite. The two on-board instruments (WFM and LAD), together with the main structural parts are indicated. **Right:** the effective area of the LAD compared to that of other launched and planned X-ray missions.

Studies carried out during the assessment phase of LOFT proved that only a 10 $m^2$-class instrument with good spectral resolution is able to successfully exploit all the relevant diagnostics of high-time-resolution X-ray spectral observations of compact objects, which provide direct access to strong-field gravity, the equation of state of ultra-dense matter and black hole (BH) mass and spin.

## 2 LOFT - The Large Observatory For X-ray Timing

A sketch of the LOFT baseline configuration is presented in Fig. 1 (left). The LOFT payload comprises two instruments: the Large Area Detector (LAD) and the Wide Field Monitor (WFM). The LAD [2] is a collimated experiment, *i.e.* a non-imaging pointing instrument with a Field of View (FoV) of ~1° and an energy resolution of ~240 eV. The LAD achieves an enormous effective area (>15 times larger than that of any previously flown X-ray experiment) by making use of 2016 Silicon Drift Detectors (SDDs) placed on 6 deployable panels. At 10 $m^2$ area, count rates on a 1 Crab source are 200000-240000 c/s, and the system can deal with sources of at least 15 Crab. The detectors, each providing a collecting area of about 75 $cm^2$, have been built on the heritage of the ALICE experiment at CERN and their design has been optimized for the detection of X-ray photons in the LOFT energy range (mainly 2-50 keV). They provide a spectral resolution of 200 eV for 40% of the photons and 260 eV for the remaining ones. Thanks to the thin detectors, particle background is much less of a concern than in gas-filled counter based instruments; instead the background is dominated by Cosmic X-ray background and Earth albedo. This leads to an excellent sensitivity of 0.1 mCrab in 100 s (5σ). To constrain the FoV of the SDDs to within ~1°, the LAD makes use of micro-channel plate (MCP) collimators successfully used previously on EXOSAT and Chandra, and now under study at ESA to be employed in the MIXS instrument on-board BepiColombo.

The WFM [3] is a coded mask instrument, placed on the top of the satellite's optical bench and providing a means to continuously monitor more than 1/3 of the X-ray sky at once. The WFM comprises in total 10 cameras, each equipped with 4 SDDs and its own coded mask. Each camera has a FoV of 45°x45° and is coupled with a second orthogonal camera in order to have 1 unit with full 2D imaging capabilities. The total FoV achieved instantaneously with 5 units covers 4.1 steradian. This wide field instrument provides a huge leap forward with respect to any other X-ray monitor flown so far. It is not only endowed with the largest ever FoV, but also provides for each source in this FoV energy spectra with good resolution (~300-500 eV) in a broad energy range (2-50 keV), and event files with fine time resolution (10 μs). The WFM is also equipped with an on-board alert system (the LOFT Burst alert System, LBAS) that is able to localize any bright event in the WFM FoV and broadcast their position (~1 arcmin) and trigger time to the ground within 30s (see §3).

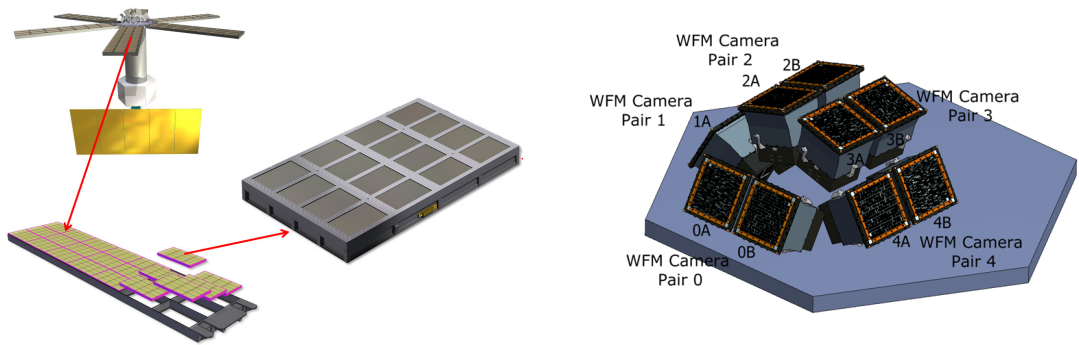

**Figure 2** Left: the modularity concept of the LAD instrument. 2016 SDDs are organized in 6 panels, each hosting 21 modules. Each module comprises 16 SDDs. Right: Distribution of the 10 WFM cameras on top of the satellite optical bench (courtesy of the LAD and WFM teams).

## 3 LOFT Science Objectives

LOFT aims to answer the two questions in §1 through the following top science goals:

- ✓ Constrain the equation of state of supranuclear density matter by the measurement, using three complementary types of pulsations, of mass and radius of at least 4 neutron stars (NS) with an instrumental accuracy of 4% in mass and 3% in radius.
- ✓ Provide an independent constraint on the equation of state by filling out the accreting NS spin distribution through discovering coherent pulsations down to an amplitude of about 0.4% (2%) rms for a 100 mCrab (10 mCrab) source in a time interval of 100s, and oscillations during Type 1 bursts down to typical amplitudes of 1% (2.5%) rms in the burst tail (rise) among 35 NSs covering a range of luminosities and inclinations.
- ✓ Probe the interior structure of isolated NSs by observing seismic oscillations in Soft Gamma-ray Repeater (magnetar) intermediate flares when they occur with flux ~1000 Crab through high energy photons (> 20 keV).
- ✓ Detect strong-field general relativity (GR) effects by measuring epicyclic motions in high frequency QPOs from at least 3 BH X-ray binaries and perform comparative studies in NSs.
- ✓ Detect disk precession due to relativistic frame dragging with the Fe line variations in low frequency QPOs for 10 NSs and 5 black holes.
- ✓ Detect kHz QPOs within their coherence time, measure the waveforms and quantify the distortions due to strong field GR for 10 NSs covering different inclinations and luminosities.
- ✓ Constrain fundamental properties of stellar mass BHs and of accretion flows in strong field gravity by (a) measuring the Fe-line profile and (b) carrying out reverberation mapping and (c) tomography of 5 BHs in binaries providing spins to an accuracy of 5% of the maximum spin (a/M=1) and do comparative studies in 10 NSs.
- ✓ Constrain fundamental properties of supermassive BHs and of accretion flows in strong field gravity by (a) measuring the Fe-line profiles of 20 AGNs and for 6 AGNs (b) carry out reverberation mapping and (c) tomography, providing BH spins to an accuracy of 20% of the maximum spin (10% for fast spins) and measuring their masses with 30% accuracy.

We give a few examples below, which show how LOFT will achieve these science goals.
*The material presented here is extracted from the mission "Yellow Book", which was written thanks to the effort of a large team (comprising all LOFT science working groups, WG). The Yellow book*

*presents a more exhaustive discussion on the LOFT science case and will be made publicly available by ESA at the end of 2013.*

## 3.1 EOS Science Requirements: LOFT Dense Matter

NSs are the densest objects in the Universe, and provide a unique laboratory to study matter in extreme conditions that cannot be replicated elsewhere. In the core of NSs, matter is compressed to densities up to ten times nuclear, achieving physical conditions at which the properties of matter are still largely unknown. This is one of the great questions of modern physics. The state of matter at these extreme conditions can be efficiently probed through the reconstruction of the equation of state (EOS) at high densities. The EOS relates the pressure, density and temperature of material inside the star, and thus provides access to all key aspects of the relevant microphysics. It has been demonstrated that there exists a unique mapping between the EOS and the neutron star mass-radius relation (NS radius R as a function of mass M [4]). By making use of its unique capabilities, LOFT will provide three independent techniques to perform measurements of M and R in a large sample of NSs:

- Pulse profile modelling: it has been demonstrated that $\sim 10^6$ pulsed photons are required to reconstruct the shape of the pulse profiles and the relevant spectral parameters in accretion-powered millisecond pulsars and/or in the burst oscillations displayed during the rise and tail of thermonuclear X-ray bursts, in order to constrain the star's M and R to within an accuracy of 4% and 3%, respectively. These accuracies are required to determine uniquely the EOS. Such numbers of counts can be collected very efficiently within reasonable observing times by a 10 m$^2$-class experiment such as LOFT, and will be attained for ~10 NS selected for their favourable geometries and distribution over a range of masses and/or radii in the M-R diagram [5].

- Spin measurements: the spin distribution of NSs provides an independent constraints on their EOS. The fastest NS provides the most stringent constraint as, at the very simplest level, the limiting spin rate at which mass-shedding occurs for a rotating star is a function of its mass and radius. LOFT will have a unique sensitivity also to very short lived pulsations, and will thus be able to discover many more spinning NSs, probing in particular the highest frequency range. Using state of the art binary pulse search methods LOFT will reach unprecedented sensitivity levels to very weak pulsations in bright sources.

- Asteroseismology: with its huge effective area LOFT has a unique sensitivity to the detection and measurement of neutron star seismic vibrations. The latter, discovered by RXTE [6, 7], opened new perspectives to study neutron star interiors. LOFT will most efficiently detect these seismic oscillations during intermediate flares displayed by magnetars (occurring much more often than the rare giant flares where the oscillations have been first discovered). These events can be observed off-axis while shining through the LAD collimators at energies ~40-60 keV; the high sensitivity of the LAD permits efficient measurement of the oscillations within the required time scales and accuracy without pointing directly at the source of the flare.

Figure 3 shows how LOFT will be able to reconstruct the neutron star EOS through the measurements described above, leading to unprecedented constraints on still unexplored properties of matter at extreme conditions.

From the present-day to the mid-2020s, some effort will also be spent by other facilities in investigating "matter under extreme conditions", using NSs as probes. Radio telescopes, such as SKA, LOFAR, ASKAP and MeerKAT, will be able to provide new EOS constraints only if the maximum NS mass record is broken, alternatively, current measurements may have already reached the maximum mass. In either case, radio observations will not deliver the precision radius measurements needed to measure the EOS, which is necessary to pick out the correct EOS. At present there is only one known radio source, the Double Pulsar, whose moment of inertia (via spin-orbit coupling) will be determined to within 10% within the next 20 years [22, 23]. This would result in a constraint on R ~ 5%. A similar conclusion applies to Gravitational wave telescopes, such as Advanced LIGO and VIRGO, as they are expected to achieve uncertainties of ~10% (1σ) in R from observations of the closest detected binaries [10]. The NASA mission NICER will have the bright pulsar PSR J0437-4715 as single primary target, and will provide for this a measurement of the NS radius at ~2% accuracy (other uncertainties in mass and pulsar emission mechanism might still affect this measurement). Although this NICER data point would thus constrain the EOS, LOFT will be the only facility to reconstruct the EOS.

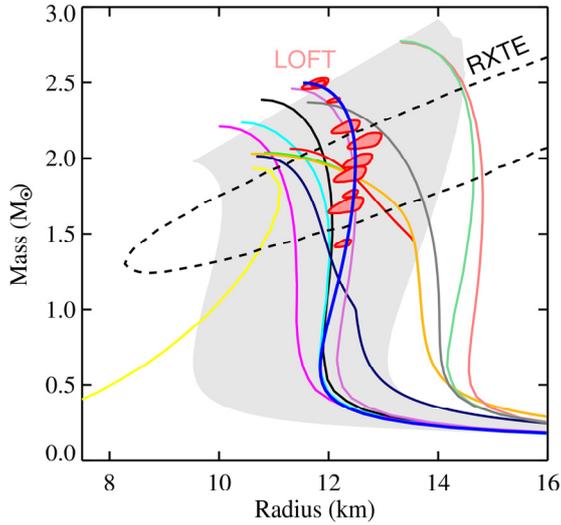

**Figure 3** Measurements of neutron star masses and radii (red error regions) that LOFT will be able to achieve by exploiting the pulse profile modelling technique summarized in §2.1. These measurements will tightly constraint the neutron star mass-radius relation and thereby the EOS of dense matter and provide a unique probe of the still unknown physical properties of matter at these extreme conditions. Mass-radius relations resulting from different EOS models are shown with solid lines which depend on both the composition of supranuclear density matter (from pure nucleonic to strange quark matter) and more generally on the nature of the strong force. Only EoS compatible with the recent discovery of two pulsars with masses ≈2M⊙ [8, 9] are shown. For comparison, a typical constraint on the EOS that could be obtained with RXTE is shown (courtesy of A. Watts and the DM WG).

## 3.2 SFG Science Requirements: LOFT Strong Field Gravity

The second set of LOFT's main science goals focus on testing for the first time the behaviour of matter in strong-field gravity, a topic that has the keen interest of both the physics and astrophysics communities. While gravitational wave detectors (such as as Ligo/Virgo) and the recently approved ESA L3 mission will be investigating GR in case of dynamic spacetimes [11], LOFT will probe the motions of matter subjected to strong-field stationary spacetimes of both weak and strong curvature. In particular, it will be able to perform direct measurements of how the motions of matter located a few gravitational radii from a black hole (BH) are affected by its strong field gravity. GR predicts these to grossly deviate from Newtonian motions and include qualitatively new effects. LOFT will be able to make use of the two most important direct diagnostics of strong-field gravity near stellar-mass BH in X-ray binaries and Active Galactic Nuclei (AGN), i.e. the relativistically broadened iron-lines and relativistic timescale variability (in particular the Quasi Periodic Oscillations, QPOs; [12]). The LOFT capabilities will permit for the first time the study of the variability of relativistic lines on timescales typical of regions dominated by the strong field gravity. Figure 4 shows that LOFT will easily resolve with unprecedented signal-to-noise ratio (S/N) changes in the iron line profiles expected to occur due to the variable disk illumination of the hot flow around a BH that is precessing as a consequence of the relativistic frame dragging effect [13]. This quasi-periodic precessional motion gives rise to the 0.03-30 Hz QPOs observed from accreting stellar mass BH and NS systems.

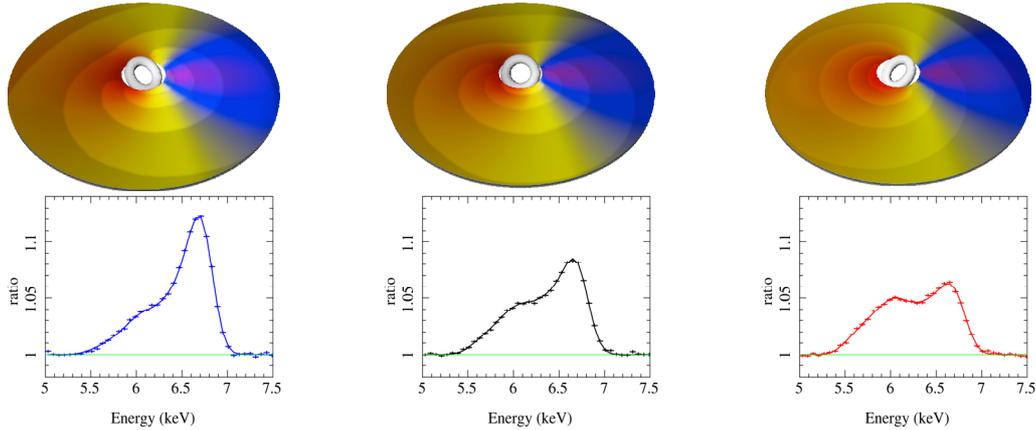

**Figure 4** LOFT detection of variable iron line profiles during the rise (left), peak (middle) and fall (right) phase of the QPO generated by the precession of the inner flow around the BH due to the relativistic frame dragging effect (courtesy of A. Ingram, M. van der Klis and the SFG WG).

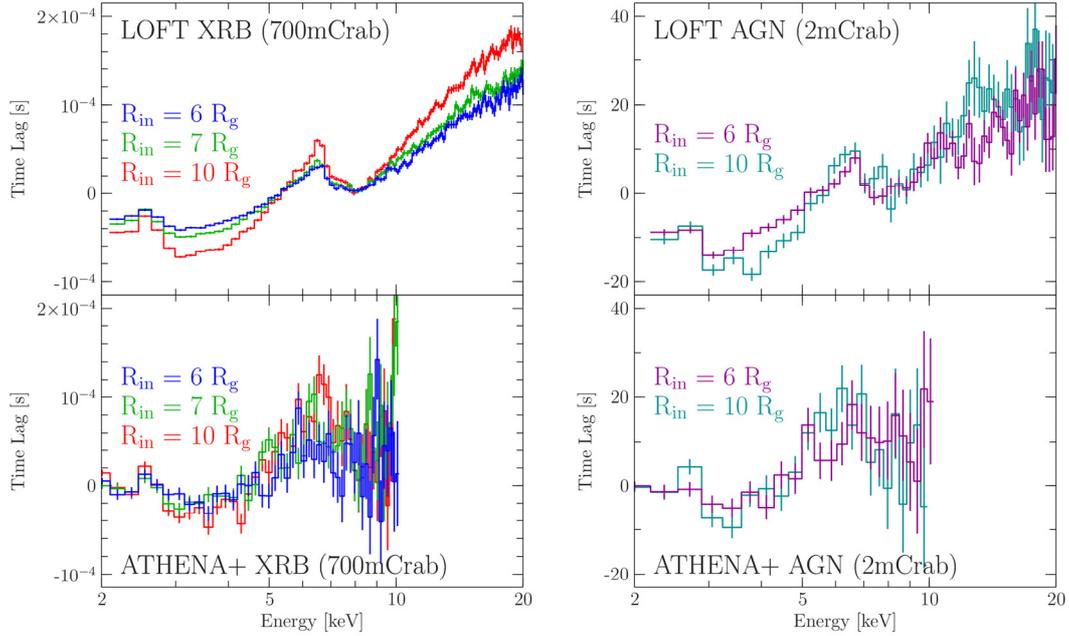

**Figure 5** Time lag spectra due to reverberation light travel time delays using the 60-200 Hz broad-band variability in a 10 $M_\odot$ XRB BH (left) and the 0.3-3mHz variability in a $4 \times 10^6$ $M_\odot$ AGN (right). Three different inner disk radii are illustrated for the XRB, two for the AGN. Disk inclination is 30°. For comparison the results that would have been obtained with Athena+ are shown. XMM-Newton would have achieved slightly worse results (courtesy of P. Uttley and the SFG WG).

LOFT will also provide breakthrough measurements of time lags expected as a consequence of "reverberation" in BH systems. Indeed it is well known that the variability of the incident hard X-ray flux from, e.g., the hot flow around a BH onto the disk generates light travel time lags between different energy bands (Fig. 5). As these reverberation measurements encode the various relativistic effects distorting the line profiles, including redshift and strong-field Shapiro delays, as a function of absolute radius [14], BH mass and spin can be directly derived from the measured time lag spectra, and the GR predictions of the radial dependencies verified at high precision (both in XRBs and AGNs). LOFT will also be able to track quasi periodic distortions in the spectrum of XRBs due to orbiting hot spots in the disk, especially close to the energies of the iron lines. An example is shown in Fig. 6.

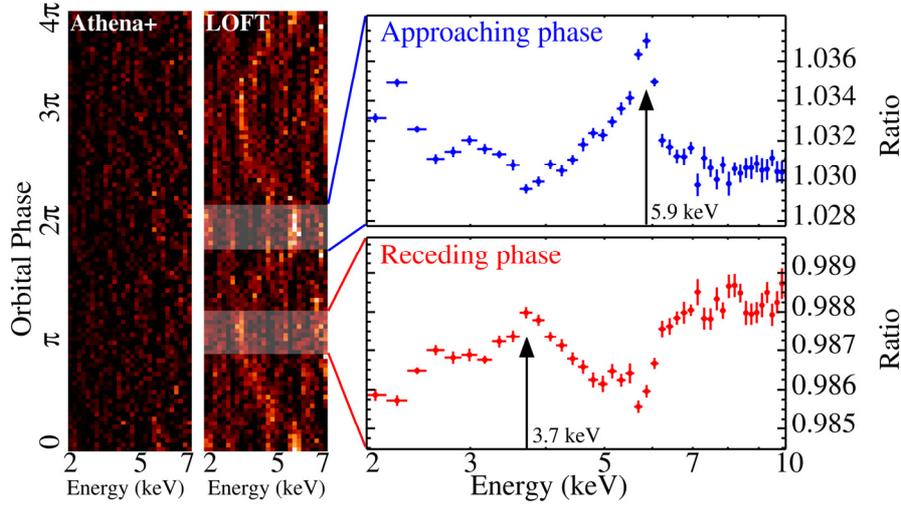

**Figure 6** Simulation of a ~216 Hz 2% rms amplitude QPO produced by patterns of hotspots (about 1 gravitational radius $r_g$ wide) in the disk (at 6 $r_g$) occurring randomly in time and azimuth, and orbiting at frequencies near the QPO frequency. The periodic curve in the dynamic energy spectrum (left) is the Fe line shifting up and down in energy due to the quasi-periodic Doppler shifts caused by the orbiting patterns, reconstructed using Fourier methods. This radial velocity curve for orbital motion in strong field gravity is non-sinusoidal [16]: the rise to maximum line energy and the maximum itself takes less time than the fall to minimum and the minimum itself. LOFT measurements can be seen to clearly exhibit these GR-predicted effects in strong-field orbital motion. For the simulation a total exposure time of 100 ks of a 1 Crab BH transient in the intermediate state is assumed. A quality factor Q=5 is assumed for the QPO, consistent with observations [17]. Two cycles are shown. Image was processed for display purposes to remove steep gradients due to Doppler boosting (courtesy of P. Uttley and the SFG WG).

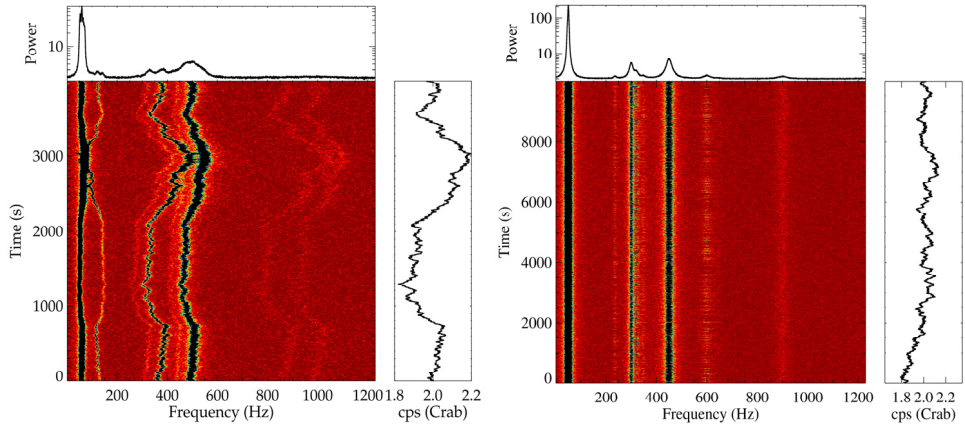

**Figure 7** Dynamical (time-frequency) power spectra of QPOs in a 1 Crab BH of 7.1 $M_\odot$ and spin a* = 0.60 at disk inclination of 63° observed with LOFT for two different epicyclic-frequency models. The tracks in the images show the QPO frequency variation (or the lack of it) in either model. Left: Relativistic precession model [18]. Frequencies observed due to the emission of an elongated (315°) luminous region with radial extent of 0.5$r_g$, in geodetic motion at radii that vary with flux (right panels) between 4.0 and 4.5 $r_g$. Frequencies most clearly visible (low to high frequency) are due to: Lense-Thirring precession, radial epicyclic motion (note the inverse dependence on radius compared to all other frequencies), periastron precession, orbital motion; several other combination frequencies are present as well. Right: Epicyclic resonance model [19]. Radial and vertical epicyclic frequencies and their harmonics are observed due to vibrations of a toroidal mass flow in 2:3 resonance

at 300 and 450 Hz. The low frequency QPO is not predicted in this model and arises through another mechanism (courtesy of D. Barret, K. Goluchova, Bakala, L. Stella, G. Torok and the SFG WG).

LOFT will also be capable of providing exceptional results on the measurements of the average line profiles both in XRB and AGNs. This will permit us to use state of the art reflection models to measure the BH spin in both type of systems. Because binary accretion roughly preserves spin and hence the birth record [15], stellar-mass BH spins carry information on BH formation in stellar collapse. Conversely, the spins of supermassive BH directly reflect their growth history through capture of gas and stars, and through mergers. The spin is thus a key parameters to understand the formation and evolutions of BHs. Simulations performed by the LOFT science working group demonstrate that the spins and inclination angles of BHs can be measured easily from the average line profiles to within the accuracies mentioned in the relevant science requirements by using exposure times as short as 100 s in XRBs and 10 ks in AGNs.

As a further probe of strong GR effects, LOFT will be able to measure BH high frequency QPOs (currently barely detected) at enormously high S/N, in some cases coherently, and hence will be able to distinguish easily between different models proposed to interpret these features, measuring the orbital, epicyclic and precessional motions of the plasma in the strong field region (Fig. 7).

## 4 LOFT Observatory Science

Since many of the sources that will be observed to achieve the LOFT top science goals are transient, an imaging WFM with an unprecedentedly large FoV has been designed to monitor large portions of the sky at once (>1/3) and provide interesting targets to be followed-up with the LAD. The exceptional capabilities of the WFM, as compared to previous X-ray monitors, make LOFT a unique discovery machine for the variable and transient sky, and thus also a fundamental complement to other "time domain astronomy" facilities that will be operational in the 2020's (such as LSST and SKA). Figure 8 provides a quick glance into the anticipated WFM capabilities, but previous experience with large sky monitors confirm that unforeseen types of source/events can be expected every time these instruments achieve a significant leap in their performance. In this context the WFM is not only unique for its incredibly large FoV, but also because it provides high time (~10 μs) and fine spectral resolution (~300-500 eV) simultaneously for any source in this FoV. The WFM is also endowed with an on-board system (the LBAS, see §2) that allows the detection of the trigger time and position of bright impulsive events (e.g., Gamma-ray Bursts, GRBs) and broadcast this information to the ground within 30 s (to this aim a system of about 15 VHF receivers will be placed around the Earth equator to cover the LOFT orbit and receive alert messages from the spacecraft). It will thus be possible to study efficiently the prompt emission of bright impulsive events in a broad energy range (2-50 keV) and quickly trigger follow-up campaigns with the LAD and other facilities that will be operating in the 2020s. In particular, observations with the LAD will enable deep spectroscopic and time variability studies across objects exhibiting a wide range of highly energetic physical phenomena powered by accretion, thermonuclear and magnetic processes as well as various types of gravitational collapse. For objects ≥1-2mCrab it will provide the best variability information and mid-resolution spectroscopy observations ever across the 2-30 keV band.

Given the incredible leap in effective area, FoV, and sensitivity of the instruments on-board LOFT, the range of astrophysical problems that will receive new stimuli from the mission's capabilities is enormous and cannot be summarized here. A more detailed discussion can be found in the mission Yellow Book (to be published by ESA) and in the large number of scientific papers mentioning the breakthrough capabilities of LOFT that can be found through the Astrophysics Data System (ADS). At the time of writing about 140 papers are listed in ADS, ranging from accreting systems (with BHs, NSs, and white dwarfs) to strongly magnetized objects, GRBs, Terrestrial Gamma-ray Flashes, tidal disruption events, flaring stars and cosmology.

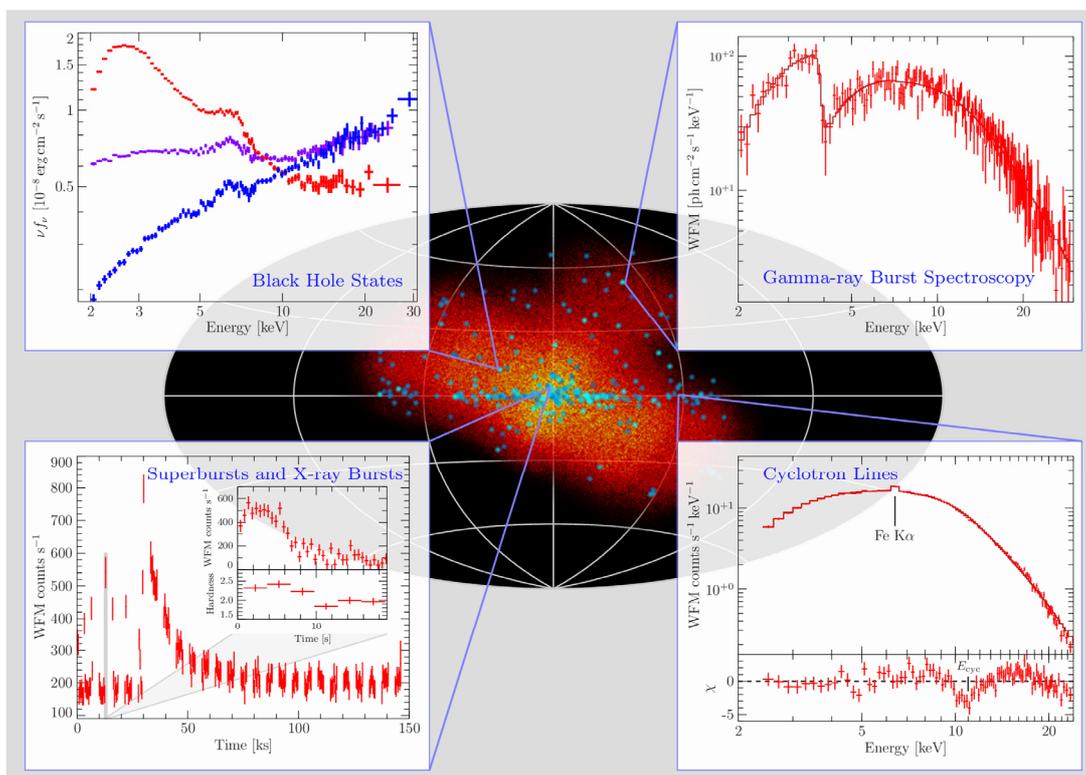

**Figure 8** Capabilities of the WFM instrument. Some astrophysical events that can be detected during pointings performed in the direction of the Galactic Center are represented. As an example, it is shown how the WFM would be able to: (i) track state changes during a transient BH outburst and detect the relativistic Fe Kα line (top left panel, 15 ks exposure, source flux 300 mCrab); (ii) measure the energy and follow the evolution of cyclotron features during the outbursts of neutron star in X-ray pulsar binaries, unveiling the strength and geometry of their magnetic fields close to the polar caps (bottom right panel, 20 ks exposure, assumed line energy of ∼ 11 keV and ∼ 22 keV; [20]); (iii) detect gamma-ray bursts and reveal transient absorption features in their prompt X-ray emission (top right panel, here a feature similar to that detected in GRB990705 was simulated; [21]); (iv) catch bursting activity and rare events such as superbursts (bottom left) from known and newly discovered sources (the WFM FoV is represented in orange in the image in Galactic coordinates; courtesy of J. Wilms, C. Schmid and the OBS WG).

## 5 Monitoring the X-ray sky

We emphasised above that many of the most interesting sources that LOFT will observe are transients and/or undergo changes in their spectral/timing states that need to be monitored carefully in order to schedule observations at the right moment. For this reason, the ground segment and science operation strategy of LOFT have been designed in order to maximize the science return of the mission and the exploitation of the LOFT data products by the science community at large. Both the data from the LAD and the WFM will be inspected as soon as available on the ground in order to search for transient sources and/or interesting events. A preliminary version (NRT, near real time) of the WFM data will be typically available within ~3 hours from the observation, while the LAD NRT data might require up to 1 day to be fully accessible on the ground. Consolidated data (including all necessary correction due to transmission problems that might occur in the NRT data and consolidated auxiliary files) for both instruments are expected to be available within 1 week from the observation.

All WFM data will be made publicly accessible as soon as they are downlinked from the spacecraft and processed "on the fly" on the LOFT Science Data Archive (LSDA) at ESA. Proper interfaces will be provided to access science products and monitor the X-ray sky in near real time. For the LAD data a "proprietary" period of 1 year will apply, and thus NRT data will only be distributed to the science principal investigators (PIs) in order to provide them a means to check the preliminary results of their observations and optimize the instrument set-ups for the following campaigns. The alert messages broadcasted by the LBAS through the VHF antennas will provide the world-wide community, as well as robotic telescopes and other facilities operating at the time of LOFT, with "heads-up" information to promptly follow-up bright impulsive events discovered by the WFM. A LOFT alert center will also be available to check at any time these alerts and provide updates to the community.

**WFM NRT data** (2-50 keV):
Publicly available < 3 hours
Provide for any source in the field of view:
    Sky mosaic & source identification
    Lightcurves (max time res. 10 μs)
    Spectra (energy res. <500 eV)

**LAD NRT data** (2-30 keV):
Distributed daily to the observations PI
Used to monitor on-going observations
Include for the observed source:
    Event files & lightcurves
    Power and energy spectra

**LOFT Burst Alert System**:
On-board localization of impulsive events (~1 arcmin)
Broadcast of time and position to the ground (< 30 s)
Provide "heads-up" for multi-wavelength facilities

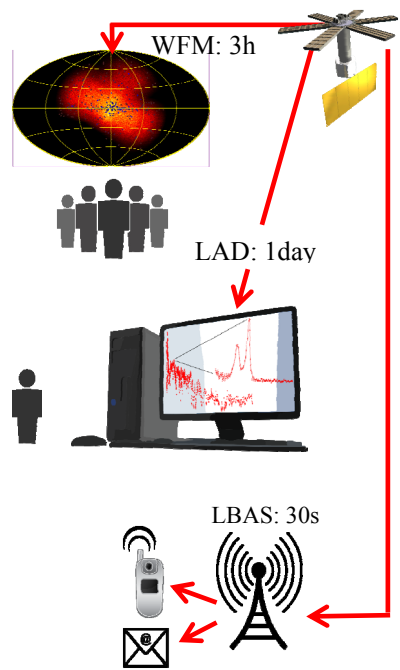

**Figure 9** Quick looking facilities provided by LOFT to efficiently monitor the X-ray sky and maximize the science return and data exploitation of the mission.

## 6 Status of the mission

LOFT completed the assessment phase study at the end of 2013, about 3 years after the initial selection in February 2011 as a candidate M3 mission. The ESA M3 mission assessment activities for LOFT included an internal ESA study and two parallel industrial studies, which were completed in early 2013. Technology development activities were carried out in parallel (and some of them are on-going) to support the advancements in the design of relevant payload items.

Mission descopings are relatively common during the assessment studies; in the case of LOFT it was instead possible to enhance the mission capabilities with respect to those foreseen in the original proposal submitted in 2011. In particular, following the selection of a launcher with greater capacity necessitated by the need to fit the required 10 m$^2$ effective area with proper margins, the following aspects were significantly improved (still keeping all resources within the envelope of a M–size ESA mission):

- The instantaneous field of view of the WFM was increased from 3.4 sr to 4.1 sr in the anti-sun direction matching the accessible part of the sky while the sensitivity was improved by adding 6 cameras to the original 4.

- The burst alert system (LBAS) was added. This allows for fast identification of GRBs and new transients, whose trigger time and sky coordinates will be dispatched to the scientific community within 30 s.
- The instantaneous accessible fraction of the sky has been increased from the original 35% to 50% (and >65% can probably be achieved).

Regarding the most critical technological items enabling the instruments on-board LOFT, full scale SDDs have already been manufactured and the prototypes performed close to the requirements. In-depth testing campaigns proved the superior robustness of these devices against any environmental effect that might affect their performances (*e.g.*, debris and particle irradiation). The design of the front-end read-out electronics of the LAD and WFM has already proven through the first prototypes that the required specifications can be achieved quite easily, and advanced prototypes will be available in early 2014. A sample collimator for LOFT was produced and tested at the end of 2013, proving that earlier designs of these devices can be improved to reach the required specifications within a reasonable amount of development time.

Thanks to all these advances, the assessment study concluded that LOFT is rapidly maturing to the necessary Technology Readiness Level, earlier than required. Being judged as a relatively low risk program, LOFT could be ready for a launch by the end of 2022.